\documentclass{bmcart}

\usepackage[utf8]{inputenc} 
\usepackage[nolist]{acronym}
\usepackage{hyperref}
\usepackage[super]{nth}
\usepackage[inline]{enumitem}
\usepackage{natbib}
\usepackage{float}
\usepackage{graphicx}
\usepackage{caption}

\startlocaldefs
\endlocaldefs

\begin{document}

\begin{frontmatter}

\begin{fmbox}
\dochead{Regular Article}


\title{The potential of Facebook advertising data for understanding flows of people from Ukraine to the European Union}


\author[
   addressref={aff1},                   
   corref={aff1},                       
   noteref={n1},                        
   email={umberto.minora@ec.europa.eu}   
]{\inits{U}\fnm{Umberto} \snm{Minora}}
\author[
   addressref={aff1},
   email={claudio.bosco@ec.europa.eu}
]{\inits{C}\fnm{Claudio} \snm{Bosco}}
\author[
   addressref={aff2},
   email={siacus@fas.harvard.edu}
]{\inits{SM}\fnm{Stefano M} \snm{Iacus}}
\author[
   addressref={aff1},
   email={saragrubanov@gmail.com}
]{\inits{S}\fnm{Sara} \snm{Grubanov-Boskovic}}
\author[
   addressref={aff1},
   email={francesco.sermi@ec.europa.eu}
]{\inits{F}\fnm{Francesco} \snm{Sermi}}
\author[
   addressref={aff1},
   email={spiros2@gmail.com}
]{\inits{S}\fnm{Spyridon} \snm{Spyratos}}


\address[id=aff1]{
  \orgdiv{Knowledge Centre on Migration and Demography},
  \orgname{European Commission - Joint Research Centre}, 
  \street{Via E. Fermi, 2749},                     %
  \postcode{I-21027},                               
  \city{Ispra (VA)},                              
  \cny{Italy}                                    
}
\address[id=aff2]{%
  \orgdiv{Institute for Quantitative Social Science},
  \orgname{Harvard University},
  \street{1737 Cambridge St, K333},
  \postcode{02138},
  \city{Cambridge (MA)},
  \cny{United States}
}


\begin{artnotes}
\note[id=n1]{Equal contributor} 
\end{artnotes}

\end{fmbox}


\begin{abstractbox}

\begin{abstract} 
This work contributes to the discussion on how innovative data can support a fast crisis response. We use operational data from Facebook to gain useful insights on where people fleeing Ukraine following the Russian invasion are likely to be displaced, focusing on the \acl{EU}. In this context, it is extremely important to anticipate where these people are moving so that local and national authorities can better manage challenges related to their reception and integration. By means of the Ukrainian-speaking Monthly Active Users estimates provided by Facebook advertising platform, we analyse the flows of people fleeing the country towards the European Union. At the fifth week since the beginning of the war, our results indicate an increase in the number of Ukrainian-speaking Facebook users in all the EU countries, with Poland registering the highest percentage share ($33\%$) of the overall increase, followed by Germany ($17\%$), and Czechia ($15\%$). We assess the reliability of prewar Facebook estimates by comparison with official statistics on the Ukrainian diaspora, finding a strong correlation between the two data sources (Pearson's $r=0.93$, $p<0.0001$). We then compare our results with data on arrivals in Poland and Hungary reported by the \acs{UNHCR}, and we observe a similarity in their trend. In conclusion, we show how Facebook advertising data could offer timely insights on international mobility during crisis, supporting initiatives aimed at providing humanitarian assistance to the displaced people, as well as local and national authorities to better manage their reception and integration.
\end{abstract}




\end{abstractbox}
%

\end{frontmatter}



\section{Introduction}

Policies related to both disaster risk management and humanitarian assistance are designed with the aim of tackling the crises and associated challenges with preventive, preparedness, response and recovery actions. When it comes to population displacement as a consequence of conflicts, natural, or man-made disasters, there are specific challenges concerning populations’ health needs, safety, and well-being. In this context, availability of data is a crucial element as it can allow for rapid risk assessment and implementation of evidence-based risk management measures, yet there is still a need to improve the collection of disaster and conflict data ((EU)2021/836).
\footnote{\href{https://eur-lex.europa.eu/legal-content/EN/TXT/HTML/?uri=CELEX:32021R0836&rid=2}{(EU)2021/836}}

In order to provide assistance in terms of provision of food, shelter, healthcare, education to the displaced population, there is a need for timely data, and, often, comparable across countries. Recent events have shown that innovative data have the potential to integrate official data relevant for disasters and conflicts. The review of Bosco et al. \cite{Boscoetal2022} has shown that innovative data can offer a great geographic and temporal granularity, often, a (near-) real time availability, and an extensive coverage suitable for more immediate international comparisons.

Focusing on the current Russian military aggression against Ukraine, this paper aims to analyse the potential of innovative data for monitoring people fleeing Ukraine, and to contribute to the discussion on how such data can support crisis response. For this purpose, we estimate the flows of people fleeing Ukraine towards the \ac{EU} using the weekly variation of Facebook Ukrainian-speaking \acp{MAU}. To validate Facebook data with official data, we compare them with Ukrainian diaspora data (\textit{i.e.} Stocks of Ukrainians living abroad) provided by the national statistical offices at national level in the \ac{EU} countries.

In this context, the role of the \ac{JRC} are \begin{enumerate*}[label=(\roman*)]
  \item to support the \ac{EU} \acp{MS} in the implementation of the Temporary Protection Directive for people fleeing Ukraine, and
  \item to assess the reliability of innovative data in support to \ac{EU} policy-making in the context of migratory crisis response.
\end{enumerate*}

The paper is structured as follows: Sec.~\ref{sec:overview} offers a brief overview of the existing literature on innovative data for crises response; Sec.~\ref{sec:context} describes the context of the data needs of the the Russian invasion of Ukraine; Sec.~\ref{sec:data_and_methods} presents the dataset used in this work and the methodology; Sec.~\ref{sec:mau} shows the presents the findings of the analyses; finally, Sec.~\ref{sec:conclusion} discusses the results and concludes.

\section{Overview of the existing literature}
\label{sec:overview}

In the context of crises requiring fast response, there is a growing body of scientific literature that draws on innovative data in order to estimate migration and population displacement as consequence of natural disasters, man-made disasters and conflicts. 

The first and most extensive usage of innovative data can be found in the literature studying the impact of natural disasters (floods, earthquakes, hurricanes, etc.) on human mobility and migration. \textit{\acp{CDR}} in specific, appear to be the most explored innovative data source for studying human mobility and migration induced by natural disasters. Several studies have employed \acp{CDR} to analyse mobility and migration caused by earthquakes in many different countries (\cite{bengtsson2011improved,lu2012predictability,Wilsonetal2016,li2019detecting,Flowminder2021}), for example, Lu et al. \cite{lu2012predictability} analysed movements of mobile phone users from Haiti before and after the 2010 earthquake, Wilson et al. \citep{Wilsonetal2016} provided detailed spatio-temporal estimates of population movements following the Gorkha earthquake in Nepal in 2015, while more recently the Flowminder Foundation \citep{Flowminder2021} estimated populating displacement caused by the 2021 Haiti earthquake. \acp{CDR} have been used to study the effect of other types of natural disasters on human mobility and migration as well. For example, Isaacman et al. \citep{Isaacman2018} used weather data and \acp{CDR} to model the impact on migration of severe drought in La Guajira, Colombia, in 2014. Also, Lu et al. \citep{LU20161} use \acp{CDR} to estimate the migration in Bangladesh in the short- (hours-week) but also in the long-term (years) period, following the 2013 Cyclone Mahasen. Overall, these studies highlight the potential of \acp{CDR} to estimate the effect of specific climate event on mobility and migration, also in small geographical areas and in short-time intervals.

Other types of innovative data have also been used for analysing the impact of natural disasters on migration and mobility, although less extensively than \acp{CDR}. For example, Rayer \citep{rayer2018} used \textit{Flight Passenger data} to estimate the effect of 2017 hurricanes on migration from Puerto Rico to Florida, while Jia et al. \citep{Jia_2020} used \textit{Facebook displacement maps} to estimate population displacement during the Mendocino Complex and Woolsey fires in California. The latter showed that Facebook displacement maps can be used to estimate trends, magnitude, and spatial clustering of population displacement in case of disasters, although a representativeness bias remains in terms of demographic composition of the Facebook's user base. 

Over the past years there has been a rise in the number of studies using innovative data also in the context of man-made disasters, especially epidemics and pandemics \citep{li2021analysis,wesolowski2014commentary}. The outbreak of the COVID-19 pandemic has stimulated a lot of new ``data for good'' initiatives with the aim of supporting the risk assessment and identification of efficient risk management measures. Some examples of such data sharing initiatives include the release of mobility data to help the crisis response (e.g. \textit{Apple Mobility Trends Reports}, \textit{Google Community Mobility Reports}, \textit{Baidu mobility data}). These data have been especially useful for gaining insights on the relationship between the population mobility and the early spread of the SARS-COV2 virus \citep{Cot2021,Snoeijer2021, Yilmazkuday2021, Hu20, Lai2020}. 

Unlike natural disaster studies, the literature on conflict-induced migration has widely drawn on the so-called conflict and political violence event data, such as the \ac{UCDP}, \ac{GDELT}, \ac{ACLED}, \ac{GTD}, etc, collected using a semiautomatic annotation of events that appear in the news. Although these are not mobility data, this type of innovative data is frequently integrated with other type of innovative or traditional data to provide insights on conflict-induced migration. Carammia et al. \citep{carammia2020forecasting}, for example, integrate operational Google trends data with GDELT data to forecast the number of asylum applications in European countries for the coming four weeks. Suleimenova et al. \citep{Suleimenova2017} integrate Bing Maps, UNHCR data with ACLED database to simulate refugee movements following conflicts in Burundi, Mali and Central African Republic.

Other types of innovative data for conflict and migration studies appear to be relatively less extensively employed. In this field, an important study of Corbane et al. \cite{corbane2016monitoring} on the effects of the Syrian conflict showed that also open-access geospatial data (\textit{Night-time satellite data} and \ac{JRC}’s \textit{\ac{GHSL}}) can be used to produce accurate and timely estimates on migration and mobility in  conflict areas. Bharti et al. \cite{bharti2015remotely} combine night-time lights satellite imagery and anonymized mobile phone CDRs to analyse the population displacement in the context of the internal political conflict in Côte d'Ivoire in 2010. Similarly, with the crowdsourcing approach, the relevant information can be mined and used to analyse migration pathways following conflicts, however with significant methodological challenges \citep{Curry2019}.

Social media is another potential innovative datasource for studying migration and conflicts. In the context of the Venezuelan crisis, Palotti et al. \cite{palotti2020} showed that Facebook advertising platform can be used to assess in real-time and at sub-national level the number of migrants and their socio-economic profiles. This paper aims to contribute to the latter strand of literature and assess how social media data, and in particular Facebook advertising data, can be used to monitor human mobility and migration during conflicts.

\section{The case study of Ukraine: context and data needs} 
\label{sec:context}
As the Russian military aggression against Ukraine continues, the number of people forced to leave their houses relentlessly increases.
In this war scenario, counting people who moves within the country, namely \acp{IDP}, and those who have left the country in search of international protection (asylum seekers) becomes extremely difficult. 

On \nth{16} March 2022, there were almost $6.5$ million people displaced in Ukraine as a direct result of the war, according to the Protection Cluster\footnote{Global Protection Cluster \href{https://www.globalprotectioncluster.org/wp-content/uploads/Update-on-IDP-Figures-in-Ukraine-18-March.pdf}{IDP-Figures-in-Ukraine-18-March.pdf} and IOM's Ukraine Internal Displacement Report \href{https://displacement.iom.int/sites/default/files/public/reports/IOM_IDP_Estimates_UKR_16MAR2022_Round_1_full_report_v2.pdf}. Last accessed \nth{4} April 2022.}, a joint study by \ac{UNHCR}, \ac{IOM}, \ac{UN-OCHA}, and \ac{REACH} published on \nth{18} March.

The \ac{UNHCR} provides a daily update of the estimated number of people who have fled Ukraine towards the neighbouring countries after the military invasion\footnote{UNHCR Operational data Portal - Ukraine \url{https://data2.unhcr.org/en/situations/ukraine}. Last accessed \nth{4} April 2022.}. According to this source, on \nth{2} April the number of people who have left the country was almost $4.2$ million: 54\% to Poland, 14\% to Romania, 9\% to Moldova, 9\% to Hungary, 8\% to Russia, 7\% to Slovakia, and less then 1\% to Belarus.

Many \acp{NGO} have promptly responded to the humanitarian emergency by gathering funds, medicines, food, clothes, and essential goods, and by sending their staff to the ground to provide support and assistance to people in needs. On \nth{4} March, the \ac{EU} has responded, among others, with the activation of the Temporary Protection Directive\footnote{COUNCIL IMPLEMENTING DECISION (EU) 2022/382 of 4 March 2022 \url{https://eur-lex.europa.eu/legal-content/EN/TXT/?uri=uriserv\%3AOJ.L_.2022.071.01.0001.01.ENG&toc=OJ\%3AL\%3A2022\%3A071\%3ATOC}. Last accessed \nth{4} April 2022.}. 

Besides defining the decision-making procedure needed to trigger, extend, or end temporary protection, the Directive lists the rights for the beneficiaries of temporary protection. Among these rights, there is the access to employment, suitable accommodation, social welfare, medical care, and education for persons under 18.

\acp{MS} need to prepare quickly and be ready to host hundreds of thousands of people avoiding them to fall into a limbo made of administrative delays and logistic unpreparedness. Moreover, the \ac{EC} is looking for a fair way to financially support each \ac{MS} in its effort to welcome and accommodate people fleeing Ukraine. Therefore, it is crucial to assess the number of people reaching each of the EU countries and regions.

Unfortunately, \ac{UNHCR} warns that \textit{``data of arrivals in Schengen countries (Hungary, Poland, Slovakia) bordering Ukraine therefore only represents border crossings into that country, but we estimate that a large number of people have moved onwards to other countries''}. In other words, it provides estimates on the number of border-crossings from Ukraine to the neighbouring countries, rather than on the number of people hosted by each country. To date, there are no reliable data on the actual number of people who have left Ukraine and reached the \ac{EU} countries. This is where non-traditional data sources can help filling a gap by providing such estimates on a (near) real-time basis.

\section{Facebook advertising data}
\label{sec:data_and_methods}

This section describes data and methodology used for the analysis. Facebook's advertising platform provides anonymous and aggregated information on Facebook users through a dedicated \ac{API}\citep{zagheni2017}. It can be used to retrieve the estimates of the \acp{MAU} who are eligible to be shown an advertisement given a set of user characteristics. \acp{MAU} include users active within the 30 days prior to today. In this work, we focus on two main user characteristics, namely the country of residence and the language of the users. This latter attribute is provided by Facebook advertising platform to target people with language other than common language for a location. Since Facebook does not directly provide information on the nationality of its users, we use the language as a proxy to infer users of Ukrainian nationality. To test this hypothesis, in Sec.~\ref{sec:mau} we compare the Ukrainian-speaking \acp{MAU} relative to the month before the Russian invasion of Ukraine with official Ukrainian diaspora data in \ac{EU} provided by the national statistical offices at national level. Our implicit assumption is that the number of Facebook \acp{MAU} relative to Ukrainian-speaking users in each \ac{EU} country is fairly stable before the war and it is therefore comparable with the latest diaspora records.

It is worth highlighting that self-declared Ukrainian-speaking \acp{MAU} do not reflect the total Ukrainian population, the two main reasons being
\begin{enumerate*}[label=(\roman*)]
  \item not all Ukrainians use Facebook (in particular, under 13 people cannot open an account);
  \item Ukrainian is the language spoken by the vast majority of people in the country, but other languages are also common, in particular Russian\footnote{According to the 2001 Ukrainian census, Ukrainian and Russian are respectively the mother tongue of  67.5\% and 29.6\% of the population.  \url{http://2001.ukrcensus.gov.ua/eng/results/general/language/}}
\end{enumerate*}.
Nevertheless, Ukrainian language is not very diffuse outside Ukraine and the neighbouring countries, and its diffusion in Europe is very limited\footnote{It is estimated that 91\% of native speakers of Ukrainian live in Ukraine \citep{ethnologue}.}.
Moreover, we acknowledge that not all people fleeing Ukraine are Ukrainian nationals. In fact, the \ac{EU} Temporary Protection Directive is directed to everyone fleeing the country, regardless their nationality.

Recent studies have focused on the reliability of the socio-demographic information provided by Facebook's advertising platform \citep{grow2021reliable,sances_2021,Potzschke2017,zagheni2017}. Sances \citep{sances_2021} and Grow et al. \citep{grow2021reliable} observe that the information reported by the users themselves upon creating their account, in particular those that are less likely to change over time (\textit{e.g.} gender, age), are generally accurate and more reliable than other information which are inferred by Facebook advertising algorithms, such as the region of residence.
Grow et al. \citep{grow2021reliable} report that misclassifications between the actual characteristics of the users and the ones provided by Facebook are most likely to occur for the region of residence, which is partially inferred by Facebook and may change frequently, thereby increasing the chance for erroneous classifications. However, Sances \citep{sances_2021} states that classifications on the region of residence are more likely to be correct in larger regions than in smaller regions. Since we are looking at changes in \acp{MAU} at national scale, we assume the considered geographical regions are sufficiently large to neglect major classification errors.

It is important to highlight that Facebook estimates are not designed to match population, census estimates, or other sources\footnote{\url{https://www.facebook.com/business/help/1665333080167380?id=176276233019487}. Last accessed \nth{4} April 2022.}, and may differ depending on factors such as:

\begin{itemize}
    \item how many Facebook apps and services accounts a person has.
    \item how many temporary visitors are in a particular geographic location at a given time.
    \item Facebook user-reported demographics.
\end{itemize}

However, recent studies indicate that despite measurement issues and selection bias, it is potentially feasible to derive robust estimates of demographic indicators from tabulations of Facebook users \citep{zagheni2017,spyratos2019,Ribeiro2020,palotti2020}. The same works present approaches to generate bias-adjusted population estimates and demographic counts to derive the actual distributions for specific audiences of interest. Following the methodology proposed by \cite{palotti2020}, we use the Facebook penetration rate of the total resident population in each hosting \ac{EU} country\footnote{Facebook penetration rate as reported by \cite{Group2009}} as a correction factor for Facebook audience estimates (see Eq.~\ref{eq:3}), assuming people that have fled Ukraine are as likely to be Facebook users as the hosting-country's population. By doing so, we up-adjust the estimates correcting for the fact that not all the Ukrainians that have left the country are on Facebook.

\begin{equation}
\label{eq:3}
    \textrm{MAU}adj_{c} = \textrm{MAU}_{c}\:/\:\textrm{FB}pr_{c}
\end{equation}

In Eq.~\ref{eq:3}, $\textrm{MAU}_{c}$ and $\textrm{MAU}adj_{c}$ are respectively the original and the up-adjusted \acp{MAU} in the hosting country $c$, $\textrm{FB}pr_{c}$ is the Facebook Penetration rate of the hosting country.

One key aspect when using non-traditional data is the importance of validating them with reliable sources. To this date, public data on the actual flows of people fleeing Ukraine are very limited. We rely on data on refugee influx from Ukraine in neighboring countries available at the Operational Data Portal of \ac{UNHCR}. In Sec.~\ref{sec:mau} we compare the weekly change in Facebook \acp{MAU} with daily \ac{UNHCR} inflow data for the five weeks following the beginning of the war. The comparison is made only for Poland and Hungary since \ac{UNHCR} does not provide historical data for the other neighboring countries.

It is worth noting that since \ac{MAU} estimates refer to a 30 days time span, the target audience for a given country might be inflated by users transiting in a country to reach another country of destination; a user travelling in different countries will be counted as many times as the number of countries where he or she has interacted with Facebook application. As a consequence, when looking at the increase in \acp{MAU} through time it is not possible to discern how much of the change is to be attributed to Ukrainians merely transiting the country and how much to Ukrainians actually settling in. For the same reason, insights on outflows are not immediately visible, as the effect on the multiple counts would take some time to fade out.

\ac{UNHCR} data also have some caveats. First, they represent the arrivals (\textit{i.e.} inflow) of people fleeing Ukraine towards neighbouring countries, not the actual number of people displaced in a country at a given time. Second, the right to move freely within the Schengen area means there are very few border controls. The data of arrivals in Schengen countries (Hungary, Poland, Slovakia) bordering Ukraine therefore only represents border crossings into that country, but \ac{UNHCR} estimates that a large number of people have moved onwards to other countries. Nevertheless, these figures represent the only tried and tested publicly available information, and we use them to check if the trend we find in our data is confirmed or not.

\section{Estimating refugee flows at national level}
\label{sec:mau}
Since Facebook does not provide information on the nationality of its users, we use the attribute language declared by the users as a proxy to infer users of Ukrainian nationality. To test this hypothesis, we analyse the correlation between the Ukrainian-speaking Facebook \acp{MAU} and official Ukrainian diaspora in \ac{EU} provided by the national statistical offices at national level relative to the month before the Russian invasion of Ukraine.

Fig.~\ref{fig:mau_corr} shows the scatter-plots of the Ukrainian diaspora in \ac{EU} and Ukrainian-speaking \acp{MAU} (both original and adjusted with Facebook penetration rates) at national level relative to the month before the Russian invasion of Ukraine.


In both cases, we observe a high correlation with official diaspora data, with adjusted \acp{MAU} showing a slightly higher correlation coefficient (Pearson's $r=0.92$ for original (lower-bound) \acp{MAU}, and $r=0.93$ for adjusted (upper-bound) ones, $p<0.0001$). Therefore, the assumption to use Ukrainian language attribute from Facebook as a proxy of nationality seems to be reasonable (at least for the Ukrainians).

We then inspect the variation in the number of Ukrainian-speaking Facebook \acp{MAU} between the first prewar set of data and the last available period. We use the results as a proxy for the increase (decrease) of the number of Ukrainians within each \ac{EU} country.

We estimate the percentage share of increase (decrease) in Ukrainian stocks among the 27 \ac{EU} \acp{MS} in the last available week for each country $c$ as:

\begin{equation}
\label{eq:1}
    \Delta\textrm{UA}_{w_{5},c}\% = \frac{\Delta\textrm{UA}_{w_{5},c}}{\sum\limits_{i = 1}^{27} \Delta\textrm{UA}_{w_{5},c_{i}}}
\end{equation}

, where $\Delta\textrm{UA}_{w_{5}}$ is defined as:

\begin{equation}
\label{eq:2}
    \Delta\textrm{UA}_{w_{5}} = \textrm{UA}stock\cdot{\frac{\textrm{MAU}_{w_{5}}-\textrm{MAU}_{w0}}{\textrm{MAU}_{w_{0}}}}
\end{equation}

In the above equations, $w_{5}$ and $w_{0}$ represent the last available week and the baseline (\textit{i.e.} prewar) week respectively, $\textrm{UA}stock$ is the Ukrainian population provided by official statistical offices, and $\textrm{UA}$ is the ISO 3166-1 alpha-2 country code of Ukraine\footnote{\url{https://www.iso.org/iso-3166-country-codes.html}. Last accessed \nth{16} May 2022.}.

We find that $\Delta\textrm{UA}_{w_{5},c}\%$ is positive for all the \ac{EU} \acp{MS}. Fig.~\ref{fig:mau_incr_perc} shows the increase for the countries where we observe a significant change ($>2\%$).


Poland appears as the \ac{EU} country with the highest increment in the percentage share of Ukrainian-speaking Facebook \acp{MAU}, accounting for the $33\%$ of the total share in \ac{EU}. This is in line with the \ac{UNHCR} data, where Poland is reported as the country with the highest ``inflow of refugees from Ukraine'' as of \nth{4} April 2022.

Our final analysis focuses on the weekly increase of Ukrainian-speaking Facebook \acp{MAU} in the \ac{EU} \acp{MS}. Fig.~\ref{fig:mau_diff_norm} shows the normalized absolute change of \acp{MAU} for the five weeks following the start of the war (referred to as $w_{0}$). The countries shown here are the same of those in Fig.~\ref{fig:mau_incr_perc}. The boundaries of the ribbons in the figure represent the normalized original \ac{MAU} (lower bounds), and normalized adjusted \acp{MAU} (upper bound).


In all instances, the increase is steeper in the first weeks and tends to smooth over time, although this seems more evident for the countries bordering Ukraine. Poland is the country with the greatest increase of \acp{MAU} overall, followed by Czechia, and Germany. Each segment represents the increase of \acp{MAU} speaking Ukrainian with respect to the previous period. We do not observe any decrease in these countries.

It is interesting to observe the change in the slope between subsequent segments: in the future and as the situation will reach a new stationary point, we might expect this slope to decrease in the transiting countries, and increase in the final destinations, especially because the mentioned multiple count effect of Facebook \acp{MAU} would fade out through time.

In the same figure, we show the normalized absolute change of the cumulative daily arrivals to Poland and Hungary, as reported by \ac{UNHCR}. The normalization of the Facebook and \ac{UNHCR} data is done to ensure the comparability of the trends, and it is calculated as:

\begin{equation}
\label{eq:4}
    v^{'}_{i} = v_{i}\:/\:\max{(V)}
\end{equation}

In Eq.~\ref{eq:4}, $v^{'}_{i}$ and $v_{i}$ are respectively the normalized and the original $i^{th}$ value of either Facebook \acp{MAU} or \ac{UNHCR} dataset; $\max{(V)}$ is the maximum value of the series, with $V = \{v_{1},\dots,v_{N}\}$ and $N$ the number of total observations in the series. The original Facebook \acp{MAU} values are normalized using the maximum of the up-adjusted \acp{MAU} series to preserve the lower-bound.

From the results of our analysis, it seems that \ac{UNHCR} data and Facebook \acp{MAU} have similar trends. We only observe a major gap in the first week in Hungary, with \ac{UNHCR} reporting a much larger increase. The gap is then reduced in the second week by an increase in \acp{MAU}, with the rest of the period showing very similar trends. This might indicate that the increment of Ukrainian Facebook \acp{MAU} relative to the first week after the beginning of the war in Hungary was captured with a delay in the Facebook estimates, but that it was nevertheless present in the dataset.

\section{Discussion and Conclusions}
\label{sec:conclusion}

In this work, we analyse Ukrainian-speaking Facebook \acp{MAU} to monitor the flows of people fleeing Ukraine towards the \acl{EU} after the Russian invasion. Our results show that these non-traditional data can provide fast and preliminary insights on the effects of the ongoing Ukrainian crisis on the international migratory flows.

We find a strong correlation (Pearson's $r=0.93$, $p<0.0001$) between the Ukrainian-speaking \acp{MAU} in the first available week (prewar) and the official statistics of the Ukrainian diaspora in the \ac{EU} countries. This supports the assumption that the language attribute reported by the users can be considered a good proxy to infer the nationality of Ukrainian users, an information not available through the Facebook advertising platform.

In order to estimate the flows of Ukrainians in the \ac{EU} \acp{MS} following the military aggression of Ukraine, we compute the variation in the number of Ukrainian-speaking \acp{MAU} between the beginning of the war and the last available week in our dataset (the fifth week). We observe an increase in all the \ac{EU} countries, with Poland registering the highest percentage share ($33\%$ of the overall increase), followed by Germany ($17\%$), and Czechia ($15\%$).

By looking at the weekly absolute change of \acp{MAU} in the five weeks period following the start of the war, we show the increase in the \ac{EU} \acp{MS} by time, with Poland being the country with the greatest increase in absolute terms, followed by Czechia, and Germany. We observe a steeper increase in Facebook \acp{MAU} in all countries in the first weeks, especially in those bordering Ukraine. This might indicate that some of the users who fled Ukraine towards the \ac{EU} did not settled in the first country they reached, but most probably moved on towards farther destinations. Finally, we compare the increment curve in Poland and Hungary with data on cumulative daily arrivals from \ac{UNHCR}, and we observe that they present very similar trends.

It should be taken into account that \acp{MAU} data cover a 30 days period, and therefore data collected during the first five weeks of the war do not allow identifying possible secondary movements of people who fled the country. Depending on the magnitude of these secondary movements, there may be an overestimation of the outflows. The activity of an Ukrainian-speaking Facebook user moving from the first receiving country to another country within the 30 days is counted twice, one for each country where the user activity was recorded. As a consequence, it is not possible to discern how much of the increase in the number of Ukrainians in a country can be attributed to refugees merely transiting the country and how much to refugees actually settling in. Additional collections of data in different time intervals could allow to estimate the extent of secondary movements. With the present data, however, we can assume that the values in the \ac{EU} \acp{MS} that are not bordering with Ukraine are less biased because they more likely represent a final destination due to their distance from the origin.

Some caveats to this study should be highlighted. The first is related to the representativeness of the Facebook's data on Ukrainian population. We estimate the Facebook penetration rate in Ukraine to be around $40\%$. This estimation is based on the number of Ukrainian-speaking \acp{MAU} in Ukraine in prewar times over the Ukrainian population above 13 years, as reported by the Ukrainian statistical office\footnote{Data from \url{http://www.ukrstat.gov.ua/}. Last accessed \nth{9} April 2022.}. The reason we are not considering the population below 13 years of age in the calculation is because they are not allowed to open a Facebook account and so they are not supposed to be covered by our dataset.

Second, our sample is restricted only to Facebook users who speak Ukrainian, so we exclude from the analysis people fleeing the country who speak other languages. The reason behind this choice of sample restriction is the impossibility to distinguish, in the receiving countries, between Facebook users that fled Ukraine and are speaking, for example, Russian language and the Facebook users migrating from other Russian-speaking countries.

Due to the imposition of the martial law in Ukraine on the \nth{24} of February 2022\footnote{\url{https://www.president.gov.ua/en/news/prezident-pidpisav-ukaz-pro-zaprovadzhennya-voyennogo-stanu-73109}. Last accessed \nth{9} April 2022}, men between 18 and 60 years are not allowed to leave the country. Therefore most of the population that has fled Ukraine should be females, minors, and the elderly. A recent report on social media diffusion in Ukraine\footnote{DataReportal \url{https://datareportal.com/reports/digital-2020-ukraine?rq=ukraine}. Last accessed \nth{20} May 2020.} shows that most Ukrainian Facebook users are females ($60.4\%$), and users interact with Facebook mainly via their phones ($95.6\%$). For these reasons, even though we cannot capture the under 13, since a big portion of the target population should be female, we might assume that the actual Facebook penetration of our target audience under these circumstances is higher than in normal conditions, and therefore the insights deriving from Facebook \acp{MAU} should be more representative of the displaced persons. Moreover, since it is very likely that they are carrying a phone with them (to communicate with their parents and friends and for other reasons), they still would have the possibility to interact with the Facebook application should they need to.

Despite the above-mentioned caveats, and in the absence of official data on the number of people displaced outside Ukraine, our study shows that data derived from social media could offer timely insights on international mobility during crisis. These data could support initiatives aimed at providing humanitarian assistance to the displaced people, as well as local and national authorities to better manage their reception and integration.
Finally, the same data could allow monitoring international flows of people where this information is often missing, due to the freedom to move across countries without strict border control, as in the Schengen area.


\section*{Declarations}
\begin{backmatter}
\section*{Availability of data and materials}
All data used in this study are openly available.
Recent Meta data, describing the same attributes as the one used in the study, are openly available from Meta, through Facebook’s Marketing \acl{API} (\url{https://developers.facebook.com/docs/marketing-apis/}). We confirm that we, as authors, did not have any special data access privileges that others would not have. Due to legal requirements regarding the publication of Meta data, the minimal data underlying the results of this study are available for academic purposes upon request. Data can be requested from the corresponding author Umberto Minora (\href{mailto:umberto.minora@ec.europa.eu}{umberto.minora@ec.europa.eu}), European Commission - Joint Research Centre (JRC) Demography, Migration and Governance Unit, TP 266, Via E.Fermi 2749, 21027 Ispra (VA), Italy.

\section*{Competing interests}
The authors declare that they have no competing interests.

\section*{Funding}
Not applicable

\section*{Authors' contributions}
The authors contributed equally to this work.


\bibliographystyle{bmc-mathphys} 
\bibliography{refbib}      




\section*{Figures}

  \begin{figure}[h!]
    \includegraphics{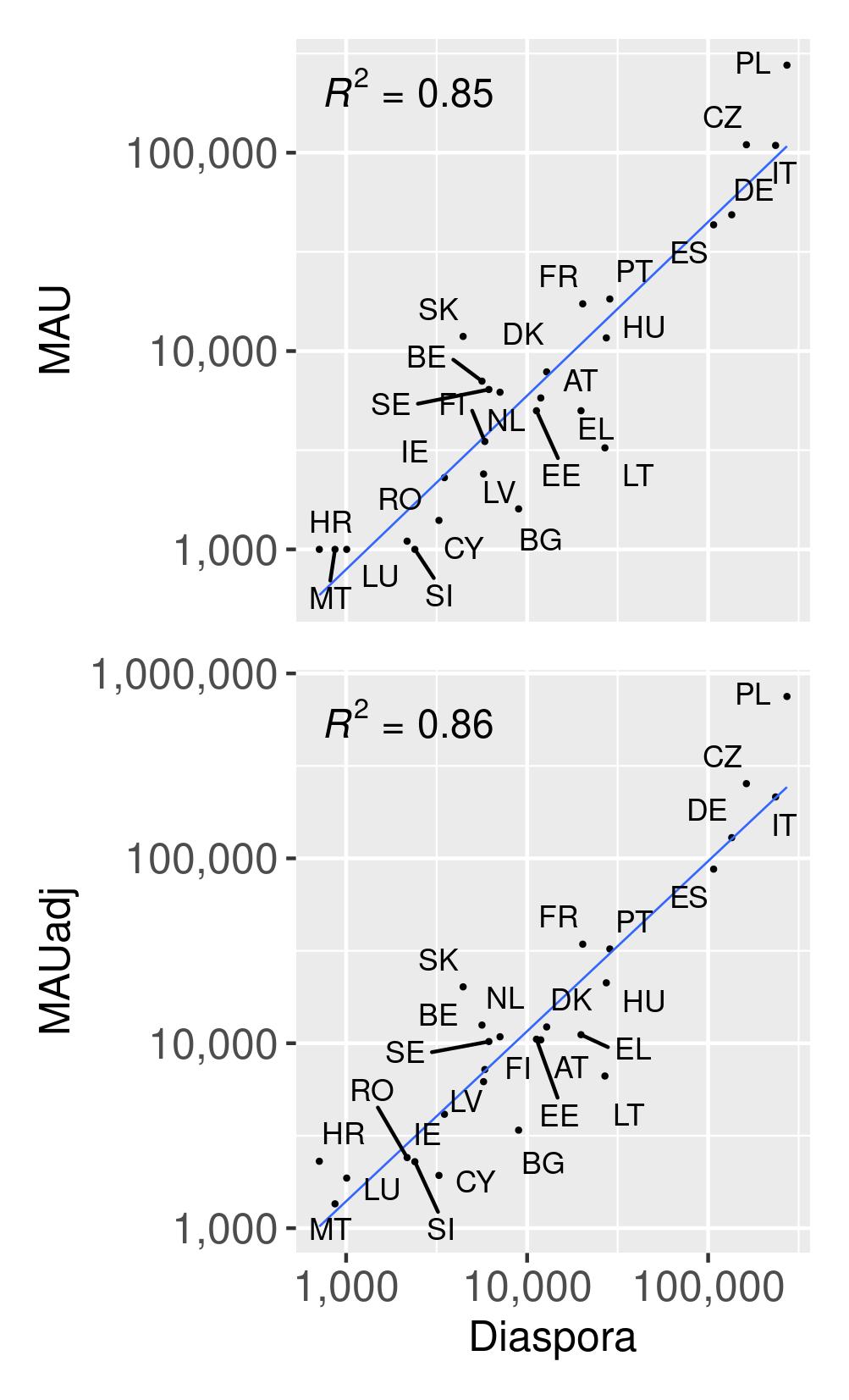}
    \caption{\csentence{Fig\_maucorr}
      Scatter-plot (logarithmic scale) of prewar Facebook \acp{MAU} at national level against Ukrainian diaspora in the EU countries (official statistics relative to 2021). Upper plot: original \acp{MAU}. Lower plot: adjusted \acp{MAU}.}
    \label{fig:mau_corr}
  \end{figure}
  
  \begin{figure}[h!]
    \includegraphics{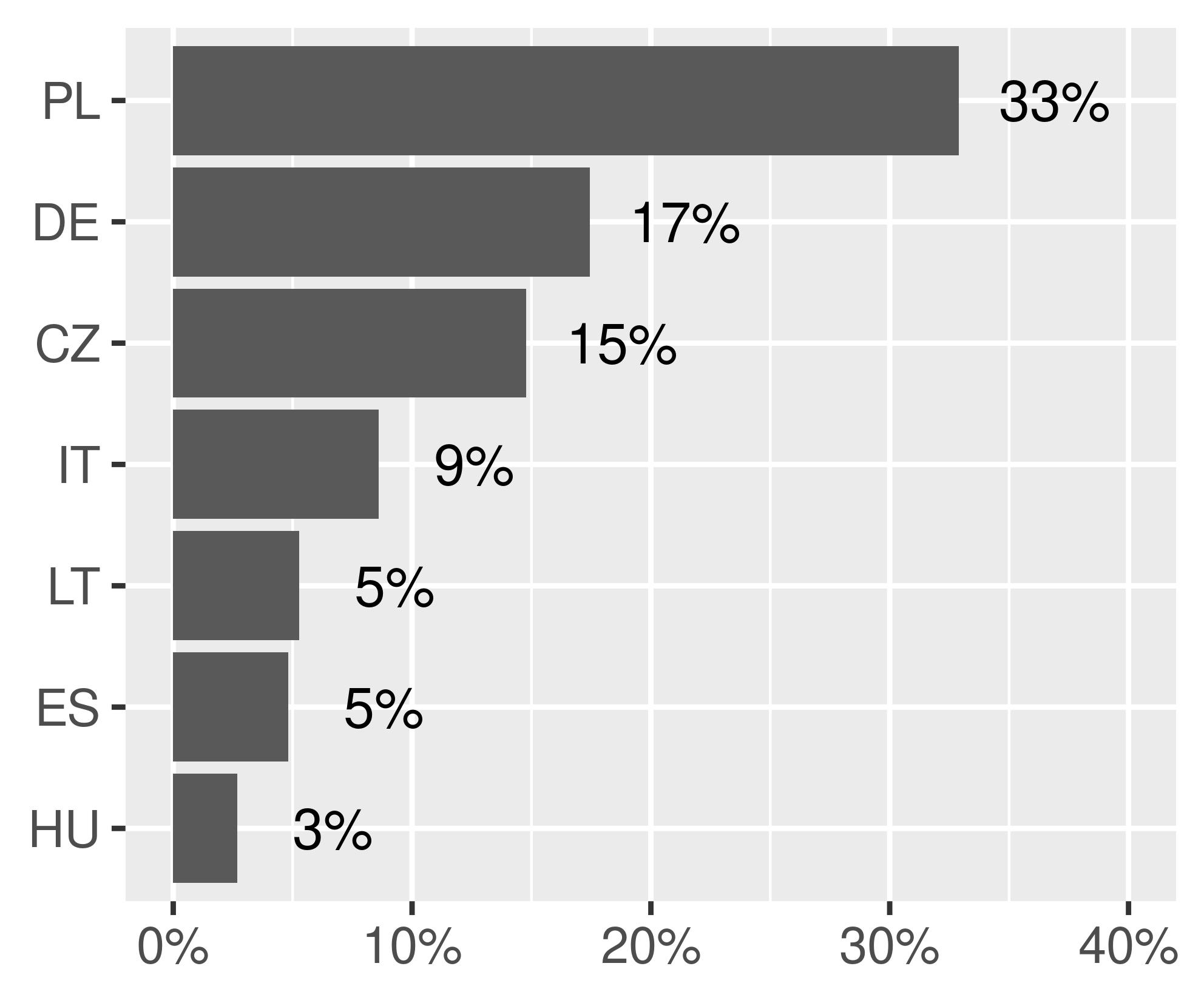}
    \caption{\csentence{Fig\_mau\_incr\_perc}
      Percentage share of Ukrainian-speaking Facebook \acp{MAU} change in the \ac{EU} countries between the beginning of the war and the fifth week. Only countries with significant increments ($>2\%$) are shown.}
    \label{fig:mau_incr_perc}
  \end{figure}

  \begin{figure}[h!]
    \includegraphics{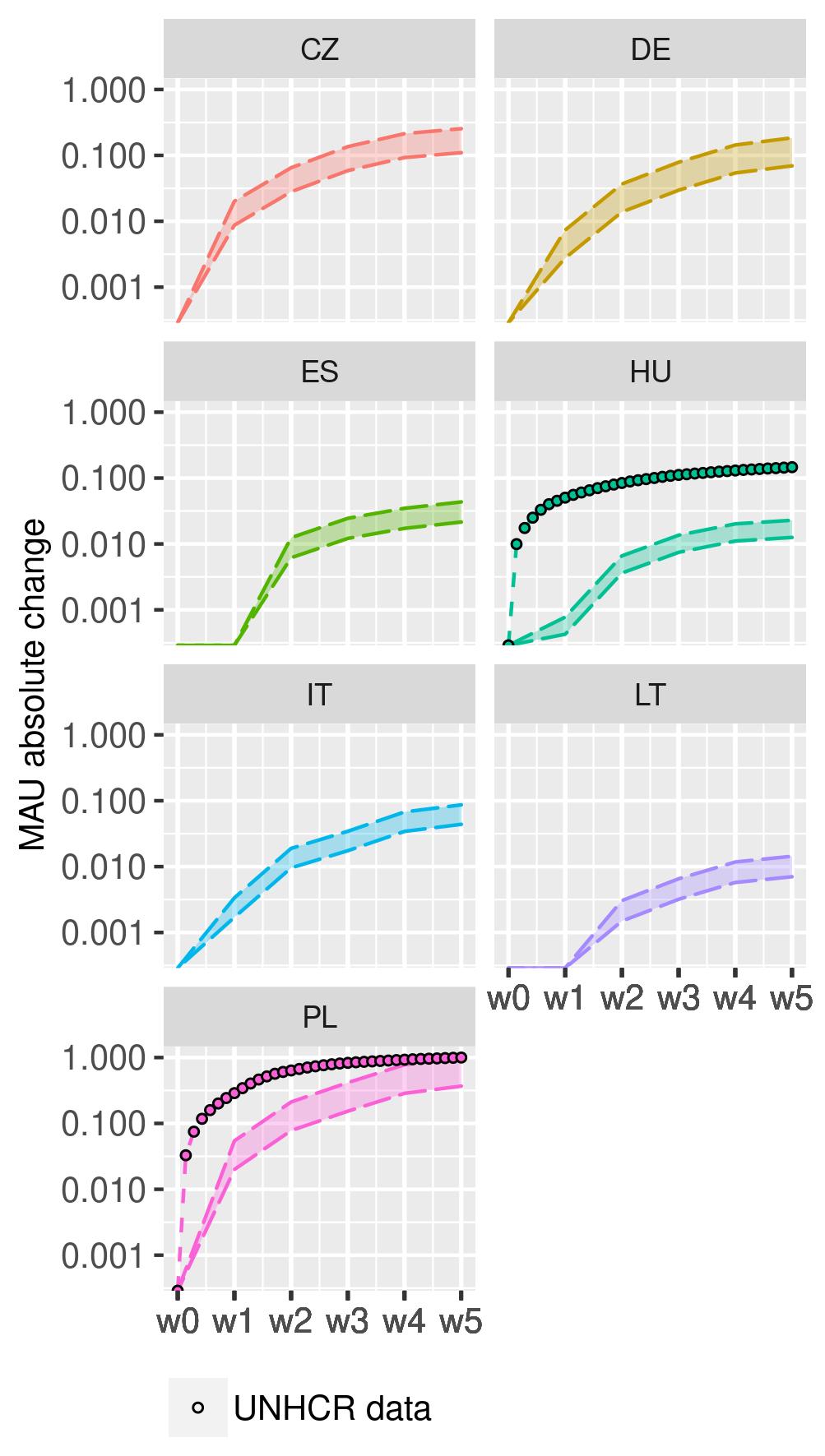}
    \caption{\csentence{Fig\_mau\_diff\_norm}
      Normalized weekly absolute change of Facebook \aclp{MAU} and normalized daily absolute change of arrivals from \ac{UNHCR} for the available weeks (logarithmic scale). $w_{0}$ represents the week when the war started.}
    \label{fig:mau_diff_norm}
  \end{figure}






\end{backmatter}

\begin{acronym}
    \acro{ACLED}{Armed Conflict Location \& Event Data Project}
    \acro{API}{Application Programming Interface}
    \acro{ANN}{Artificial Neural Network}
    \acro{AWS}{Amazon Web Services}
   \acro{BD4M}{Big Data for Migration Alliance}
   \acro{B2B}{Business to Business}
   \acro{B2C}{Business to Consumer}
   \acro{B2G}{Business to Government}
%

    \acro{CAMEO}{Conflict and Mediation Event Observations}
    \acro{CDC}{Centers for Disease Control and Prevention}
    \acro{CDR}{Call Detail Record}
    \acro{CIESIN}{Center for International Earth Science Information Network}
    \acro{CLINE}{Cline Center Historical Phoenix Event Data}
    \acro{COPDAB}{Conflict and Peace Data Bank}
    \acro{CPR}{Control Plane Record}
    \acro{data4sdgs}{The Global Partnership for Sustainable Development Data}
    \acro{D4R}{Data for Refugees}
    \acro{EASO}{European Asylum Support Office}
    \acro{EC}{European Commission}
    \acro{EFTA}{European Free Trade Association}
    \acro{EMN}{European Migration Network}
    \acro{EP}{European Parliament}
    \acro{ESOP}{European statistics on population}
    \acro{EU}{European Union}
    \acro{EVI}{Enhanced Vegetation Index}
    
    \acro{FAIL}{First Attempt In Learning}
    \acro{FRONTEX}{European Border and Coast Guard Agency}
    \acro{FDI}{Foreign direct investment}
    \acro{GBD}{Global Burden of Disease}
    \acro{GDELT}{Global Database of Events, Language, and Tone}
    \acro{GDP}{Gross Domestic Product}
    \acro{GeoSemAP}{Geospatial Semantic Array Programming}
    \acro{GHSL}{Global Human Settlement Layer}
    \acro{GIS}{Geographic Information System}
    \acro{GMDAC}{Global Migration Data Analysis Centre}
    \acro{GLH}{Google location history}
    \acro{GovLab}{Governance Lab}
    \acro{GPW}{Gridded Population of the World}
    \acro{GSMA}{Global System for Mobile Communications Association}
    \acro{GPS}{Global Positioning System}
    \acro{GTD}{The Global Terrorism Database}
    \acro{GDPR}{General Data Protection Regulation}

    \acro{H2020}{Horizon 2020}
    \acro{HFD}{Human Fertility Database}
    \acro{HMD}{Human Mortality Database}
    \acro{HRSL}{ High Resolution Settlement Layers}
    \acro{ICEWS}{Integrated Crisis Early Warning System}
    \acro{IDP}{Internally Displaced Person}
    \acro{ILO}{International Labour Organisation}
    \acro{INED}{French Institute for Demographic Studies}
    \acro{INLA}{Integrated Nested Laplace Approximations}
    \acro{INSEE}{National Institute of Statistics and Economics}
    \acro{IOM}{International Organisation on Migration}
    \acro{IoT}{Internet of Things}
    \acro{ICT}{Information and Communications Technology}
    \acro{JRC}{Joint Research Centre}
    \acro{KCMD}{Knowledge Centre on Migration and Demography}
      \acro{LIWC}{Linguistic Inquiry and Word Count}
    \acro{ML}{Machine Learning}
    \acro{MNO}{Mobile Network Operator}
    \acro{MPIDR}{Max Planck Institute for Demographic Research}
    \acro{MS}{Member State}
    \acro{MAU}{Monthly Active User}
    \acro{MS}{Member State}
    \acro{NLP}{Natural Language Processing}
    \acro{NGO}{Non-Governmental Organization}
    \acro{OEDA}{Open Event Data Alliance}
    \acro{ODI}{Open Data Institute}
    \acro{ODM}{Origin Destination Matrix}
    \acro{ODA}{Official Development Assistance}
    \acro{OECD}{Organisation for Economic Co-operation and Development}
    \acro{OSM}{OpenStreetMap}
    
    \acro{PFI}{Push Factor Index}
    \acro{PM}{Particulate Matter}

    \acro{RASIM}{Refugees, asylum seekers, immigrants and migrants}
    \acro{REACH}{Informing More Effective Humanitarian Action}
    \acro{SDG}{Sustainable Development Goal}
    \acro{SI}{Superdiversity Index}
    \acro{SNA}{Social Network Analysis}
    \acro{SNS}{Social Networking Site}
    \acro{SPEC}{Spark-based Political Event Coding}
    \acro{SRTM}{Shuttle Radar Topography Mission}
    \acro{STMF}{Short-term Mortality Fluctuations}
    \acro{SVI}{Search Volume Index}
    \acro{TFR}{Total Fertitlity Rate}
    \acro{TRE}{Trusted Research Environment}

    \acro{UCDP}{Uppsala Conflict Data Program Georeferenced Event Dataset Global}
    \acro{UK}{United Kingdom}
    \acro{UN}{United Nations}
    \acro{UNDESA}{United Nations Department of Economic and Social Affairs}
    \acro{UNECE}{United Nations Economic Commission for Europe}
    \acro{UN-CEBD}{Committee of Experts on Big Data and Data Science for Official Statistics}
    \acro{UNHCR}{United Nations High Commissioner for Refugees}
    \acro{UN-OCHA}{United Nations Office for the Coordination of Humanitarian Affairs}
    \acro{UNSD}{United Nations Statistics Division}
    \acro{US}{United States}

    \acro{VID}{Vienna Institute of Demography}
    \acro{VIIRS}{Visible Infrared Imaging Radiometer Suite}
    \acro{VLR}{Visitor Location Register}

    \acro{WEIS}{World Event/Interaction Survey}
    \acro{XDR}{eXtended Detail Record}


	
\end{acronym}

\end{document}